\documentclass[aps,pre,twocolumn,amsmath,amssymb,superscriptaddress,preprintnumbers,showpacs,floatfix,dvips]{revtex4-1}

\usepackage[dvips]{graphicx}
\usepackage{dcolumn}
\usepackage{multirow}

\usepackage{amsmath}
\usepackage{amsfonts}
\usepackage{amssymb}
\usepackage{color}

\usepackage{bm}

\newcommand{\Sec}[1]{Sec.~\ref{sec:#1}}

\newcommand{\Eq}[1]{Eq.~(\ref{eq:#1})}
\newcommand{\Eqs}[1]{Eqs.~(\ref{eq:#1})}
\newcommand{\eq}[1]{(\ref{eq:#1})}

\newcommand{\Fig}[1]{Fig.~\ref{fig:#1}}

\newcommand{\ie}{\emph{i.e.,}~}

\newcommand{\Fsmax}{\ensuremath{F_{s\mathrm{max}}}}
\newcommand{\Fcrit}{\ensuremath{F_{\mathrm{ex}}}}
\newcommand{\hatFsmax}{\ensuremath{\hat{F}_{s\mathrm{max}}}}
\newcommand{\hatFcrit}{\ensuremath{\hat{F}_{\mathrm{ex}}}}

\newcommand{\vareps}{\ensuremath{E_{\mathrm{cb}}}}
\newcommand{\scrit}{\ensuremath{s_{\mathrm{cb}}}}
\newcommand{\hatscrit}{\ensuremath{\hat{s}_{\mathrm{cb}}}}
\newcommand{\hatvareps}{\ensuremath{\hat{E}_{\mathrm{cb}}}}

\begin{document}

\title{Arrest of three-dimensional gravity-confined shear flow of wet
  granular matter}

\author{Klaus Roeller} 
\affiliation{Max-Planck-Institut f\"ur Dynamik und Selbstorganisation (MPIDS), 
  37077 G\"ottingen, Germany}

\author{Johannes Blaschke} 
\affiliation{Max-Planck-Institut f\"ur Dynamik und Selbstorganisation (MPIDS), 
  37077 G\"ottingen, Germany}
\affiliation{Fakult\"at f\"ur Physik, Georg-August-Universit\"at G\"ottingen, 37077 G\"ottingen, Germany}

\author{Stephan Herminghaus} 
\affiliation{Max-Planck-Institut f\"ur Dynamik und Selbstorganisation (MPIDS), 
  37077 G\"ottingen, Germany}
\affiliation{Fakult\"at f\"ur Physik, Georg-August-Universit\"at G\"ottingen, 37077 G\"ottingen, Germany}

\author{J\"urgen Vollmer} 
\affiliation{Max-Planck-Institut f\"ur Dynamik und Selbstorganisation (MPIDS), 
  37077 G\"ottingen, Germany}
\affiliation{Fakult\"at f\"ur Physik, Georg-August-Universit\"at G\"ottingen, 37077 G\"ottingen, Germany}

\begin{abstract}
  We study the arrest of three-dimensional flow in wet granular matter
  subject to a sinusoidal external force and a gravitational field
  confining the flow in the vertical direction.  The minimal strength
  of the external force that is required to keep the system in motion
  is determined by considering the balance of injected and dissipated
  power.  This provides a prediction whose excellent quality is
  demonstrated by a data collapse for an extensive set of event-driven
  molecular dynamics simulations where we varied the system size,
  particle number, the energy dissipated upon rupturing capillary
  bridges, and the bridge length where rupture occurs. 
  The three parameters of the theoretical prediction
  all lie within narrow margins of theoretical estimates.
\end{abstract}

\date{\today}

\pacs{
45.70.-n,
68.08.Bc,
05.70.Ln}

\maketitle

\section{Introduction}
\label{sec:Intro} 

Sudden arrest of granular flows 
is an eminent problem in
the engineering sciences \cite{MiDi2004} as well as a challenge to the
theoretical description of granular flows in a hydrodynamic setting
\cite{JaegerNagelBehringer1996,kadanoff99,SilbertErtas01,Aranson2006,JopForterrePouliquen2006,BorzsonyiEcke2007,ForterrePouliquen2008,Luding2009}. 
From the latter perspective it involves two challenges: 
(a) Appropriately incorporating the role of dissipation arising from the particle
interactions into the framework of the balance equations underlying
hydrodynamic transport equations.
And, 
(b) Addressing the role of shear stresses, of the spatial distribution of stress,
and of yield stress in systems where the flow is spatially anisotropic.

For granular systems with purely repulsive interactions recent studies
\cite{UtterBehringer2008,BerardiBarrosDouglasLosert2010,Hecke2010,TordesillasLinZhangBehringerShi2011}
put severe constraints on hydrodynamic descriptions of dense flows by
pointing out a lack of scale separation of microscopic and relevant
hydrodynamic time and length scales. Among others this gives rise to
a severe dependence of the effective material properties on the preparation history
\cite{LoisZhangMajmudarHenkesChakrabortyOHernBehringer2009}.
In contrast, hydrodynamic and continuum-mechanics considerations
appear to provide a good description for granular systems where the
hard-core collisions with restitution are augmented by (reversible)
short-range attraction between particles
\cite{TrappePrasadCipellettiSegreWeitz2001,RognonRouxWolfNaaimChevoir2006,RognonRouxNaaimChevoir2008}.
Arguably this is due to the separation of connectivity and rigidity
percolation in response to attractive interactions
\cite{LoisBlawzdziewiczOHernCorey2007,LoisBlawzdziewiczOHernCorey2008}.
This idealization of the particle interactions
\cite{PitoisMoucherontChateau2000} applies as long as high-impact
particle collisions with high capillary numbers dominate the dynamics
(see~\cite{KantakHrenyaDavis2009,DonahueHrenyaDavis2010} for recent
applications).
On the other hand, recent experimental
\cite{LiaoHsiau2010,RemyKhinastGlasser2012} and numerical
\cite{RemyKhinastGlasser2012} work on slowly moving shear flow in
dense granular systems clearly underline the important impact of
dissipation due to the hysteretic formation and breaking of capillary
bridges. Rather than accounting for the finite restitution in
collisions and assuming reversible attractive forces, the present work
therefore takes a complementary point of view: We explore slow flows
in wet systems where dissipation is arising solely from the
hysteretic nature of the capillary interaction between the wetting
liquid and the particles, i.e., it is due to the formation and
rupturing of capillary bridges between particles
\cite{Herminghaus05,MitaraiNori2006}. The hard-core collisions are
elastic.

Shear forces that drive the flow can be modeled in various
forms. Experimentally studying shear forces in granular systems can be
done, for instance, by constructing two counter-rotating cylindrical
walls (see \cite{LiaoHsiau2010} and references therein), by exploring
a flow down an inclined plane
\cite{RahbariVollmerHerminghausBrinkmann2009}. 
Numerical models also addressed the shear flow induced by applying a
cosine force field
\cite{Schulz03,Herminghaus05,roeller2009,vollmer2010}. Similarly to
the method of images, this may be used to implement zero flow
velocity at the positions envisioned for the walls.

\begin{figure*} 
  \rule{10mm}{0mm} 
  \includegraphics[width=0.4\linewidth]{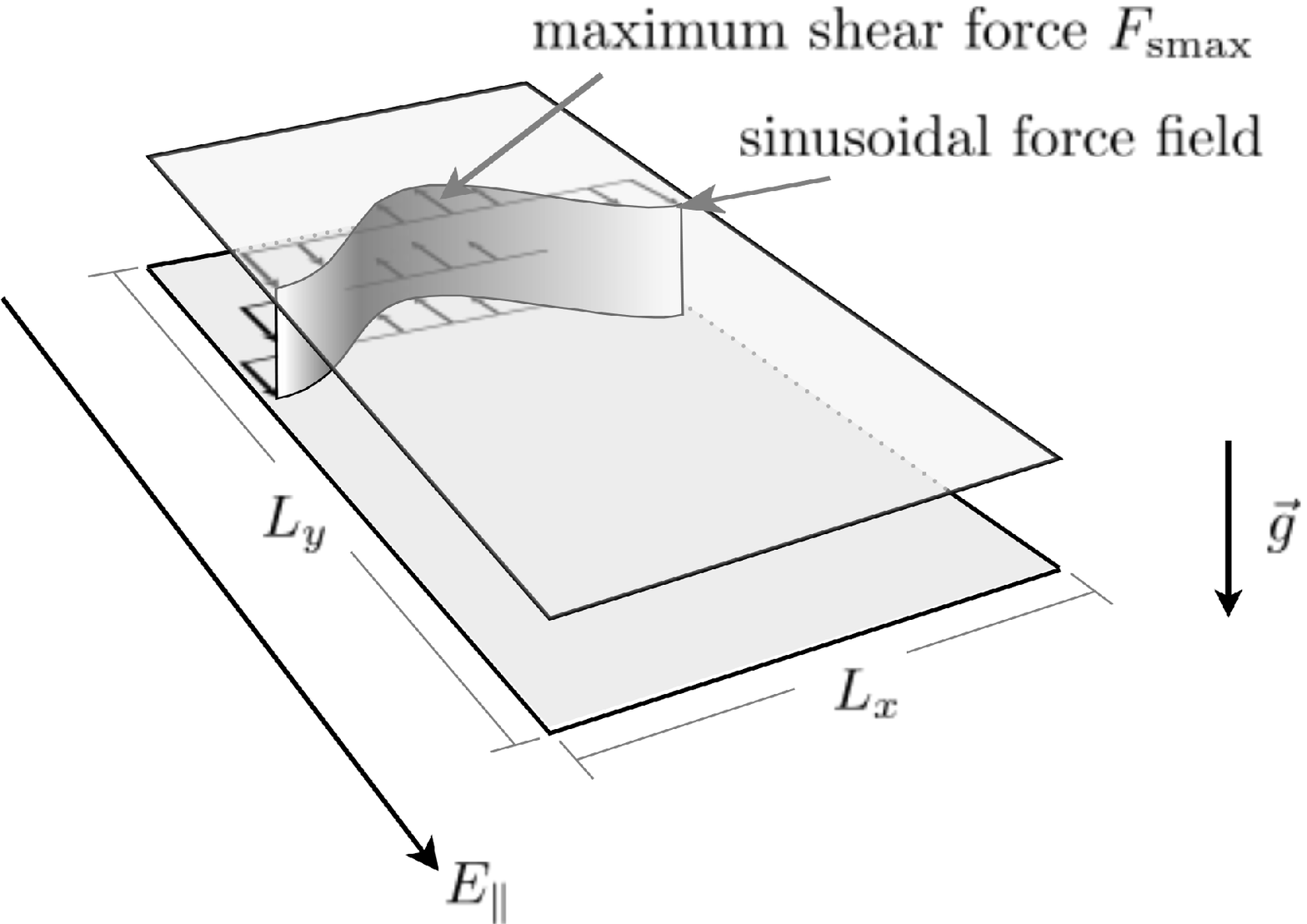}
  \rule{10mm}{0mm} 
  \includegraphics[width=.4\linewidth]{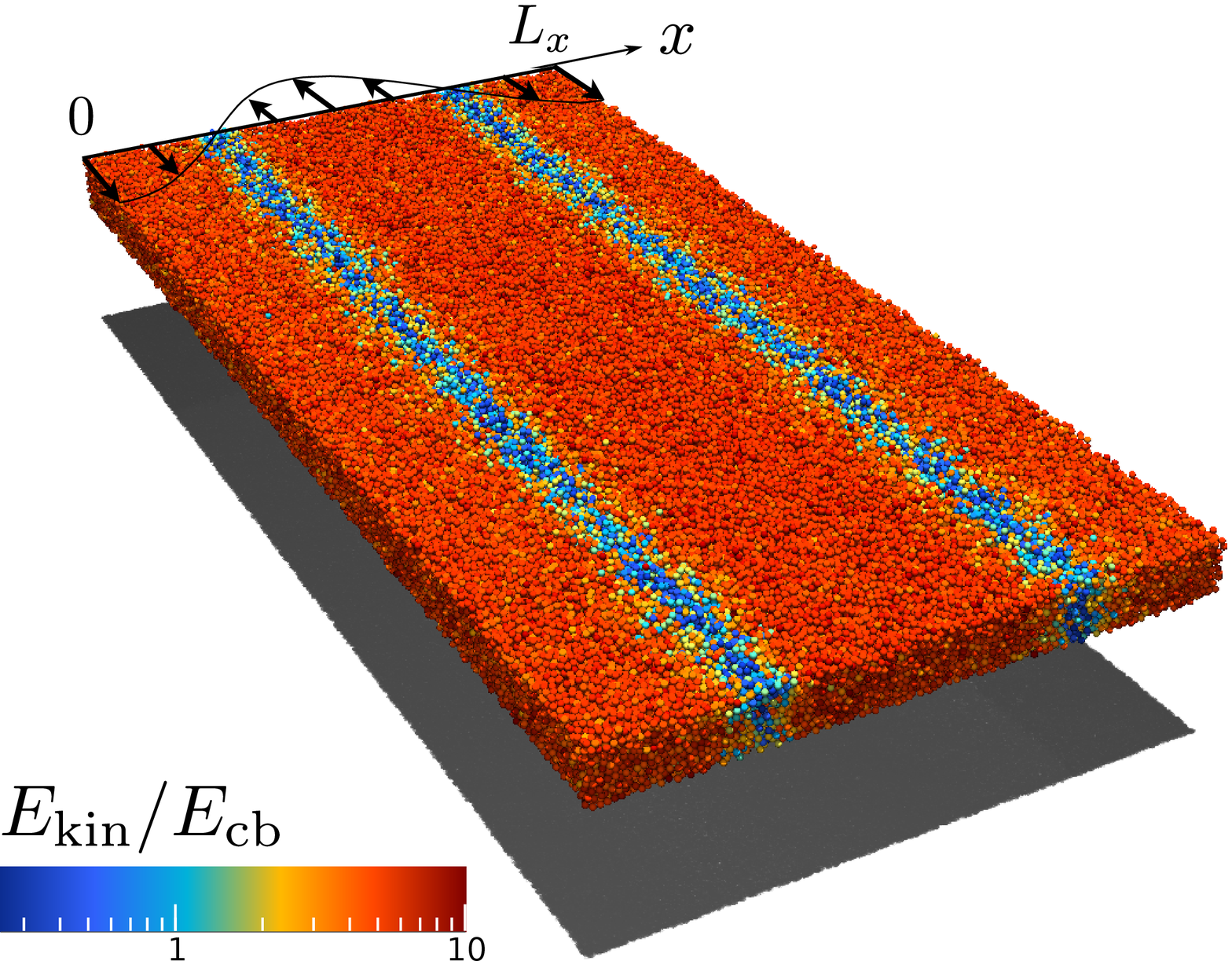}
  \hfill
  \caption{(a) Sketch of a system of width $L_x=L$ and length $L_y=2L$
    with periodic boundary conditions in the two lateral directions
    and an elastically reflecting top and bottom plate. The
    space-dependent cosine-shaped force field is indicated by the
    vertical band and by arrows.  Additionally, a gravitational force,
    $\vec g$, is acting downwards.  %
    (b) A snapshot of a simulation of $2.15\times 10^5$ monodisperse
    spheres of diameter $d$ that interact via a hysteretic square-well
    potential (see \Sec{interactions}). The cosine-shaped shear force
    field, $\vec{F}_s(x)$, is sketched on top of the figure. The color
    of each particle indicates its individual kinetic energy according
    to the color bar at the bottom left. The system size is
    ${L}=100\,d$ and ${H}=12.5\,d$, and the filling height is
    ${h}=8.8\,d$ for a shear force of amplitude $\Fsmax
    =40\,\vareps/L$, i.e., $\hatFsmax = 0.15$.
  \label{fig:EinleitungKolmogorovSystem}
  \label{fig:Snapshot}
}
\end{figure*}

Here, we will focus on the arrest of flow when the force 
driving the flow falls below a threshold value \Fcrit. 
In~\cite{vollmer2010} it was suggested that this critical force
results from the power balance between the energy-injection rate
resulting from particles motion in the external force field, and the
dissipation rate accounting for the rupture of capillary bridges
between the particles.
We will show that this approach might universally model the arrest in
vastly different settings.
To this end we describe the arrest of flow of wet, hard spheres in three
dimensions where the motion in the third dimension is constrained by a
hard wall at the bottom and a gravitational field in the vertical
direction.
For external driving forces close to \Fcrit\ all particles accumulate
at the bottom of the cell such that the packing density lies only
slightly above random close packing.
In the following ways this setting is fundamentally different from
two-dimensional flows at fixed density that were considered in
Ref.~\cite{vollmer2010}:
\begin{itemize}
\item In the previously studied setting \cite{vollmer2010} the particle
  density is prescribed by hard walls. Consequently, flow requires
  long-range rearrangements when the density is fixed close to random
  close packing. In that case the passing of particles in the flow
  requires a cooperative large scale rearrangement of large portions
  of the systems which is accompanied by the breaking of a large
  number of capillary bridges.
\item In contrast, the present study focuses on gravity-confined
  granular beds of particles, where particles can pass each other via a
  slight, local expansion of the bed in the vertical direction.  Only few
  capillary bridges must be ruptured in the resulting fixed-pressure
  setting, even though the density of the bed is always very close to
  random close packing.
\end{itemize}
In spite of these considerable differences an informed implementation
of the theoretical considerations developed in \cite{vollmer2010}
provides an excellent description of arrest in both settings.

We stress that the arrest of flow which is addressed here is
conceptually different from the jamming transition
\cite{CatesWittmerBouchaudClaudin1998,TrappePrasadCipellettiSegreWeitz2001,OHernSilbertLiuNagel2003,DroccoHastingsReichhardtReichhardt2005}.
Jamming is prevented here because the system can expand in the
vertical direction where it is only bounded by gravity (see, for
example \cite{Valverde2004} and references therein).
Furthermore, the transition is also distinct from those observed in
earlier studies on cohesive granular materials, because we focus on
the role of dissipation due to the hysteretic nature of capillary
bridge ruptures, while previous studies
\cite{TrappePrasadCipellettiSegreWeitz2001,RognonRouxWolfNaaimChevoir2006,RognonRouxNaaimChevoir2008}
addressing the transition between fluidized and arrested states dealt
with conservative attractive forces. They implemented dissipation via
grain friction and a restitution coefficient smaller than one
\cite{RognonRouxWolfNaaimChevoir2006,RognonRouxNaaimChevoir2008}, or
indirectly by treating the suspending fluid of attractive colloidal
particles as an inert background
\cite{TrappePrasadCipellettiSegreWeitz2001}.

Our paper is organized as follows. 
In \Sec{model} we describe the system, our numerical approach, and the
dimensionless units adopted for modeling. Subsequently, in
\Sec{fluidisation} we work out the power injected into the systems by
the external field, and the dissipated power due to the breaking of
capillary bridges. The power balance provides a prediction for the
critical force \Fcrit. When larger forces are applied, energy
injection dominates and there is sustained flow
\cite{roeller2009}. Otherwise, the systems relaxes into an arrested
state where the capillary bridges form a static network. 
In \Sec{phase-diagram} the predicted parameter dependence of \Fcrit\
is compared to the numerical findings. Based on only three scalar
constants we can fully describe the dependence of the stability
border in the four-dimensional parameter space spanned by 
the system size $L$,
the particle number $N$, 
the energy $\vareps$ dissipated upon rupturing capillary bridges, and 
the critical bridge length \scrit\ where rupture occurs.
(Here and in the following the subscript cb refers to
\underline{c}apillary \underline{b}ridge.)
The values of the three constants are interpreted and derived from the
model in \Sec{discussion}.
Finally, in the concluding \Sec{conclusion} we summarize and interpret
our main results.

\section{The model}
\label{sec:model}

The system is confined in a rectangular cuboid of size
$L \times 2L \times H$ 
with periodic boundary conditions in $x$ and $y$ direction,
and solid, reflecting walls in the $z$ direction. 
This box contains $N$ particles whose motion is confined in the $z$
direction by a gravitational field of uniform acceleration, $\vec g =
-g \hat z$, where $\hat z$ is the unit vector along the $z$-axis (see
Fig.\,\ref{fig:EinleitungKolmogorovSystem}).  For the shear flows
studied in the present work particles never touch the upper wall of
the container due to the gravitational confinement.

\subsection{Particle interactions}
\label{sec:interactions}

In the present study we consider monodisperse spheres
of diameter $d$, in order to suppress any additional
dynamics arising from different particle sizes. After all, 
polydisperse beads, which are subject to small shearing forces,
segregate according to their size \cite{Schulz03}, 

The dynamics in the simulation is calculated using a standard
event-driven molecular-dynamics method which has been described in
detail in
Refs.~\cite{Fingerle08,HuangRoeller09,roeller2009,ulrich09,ulrich09_PRE,ThesisKlaus10}. 
For the sake of a self-contained exposition we only briefly summarize
the particle interactions. 
Following \cite{Herminghaus05} the effect of particle adhesion due to the
capillary bridges is modeled as follows
\begin{itemize}
  \item[i.] Capillary interaction gives rise to pair forces between
    particles only. 
  \item[ii.] When particles are not connected by a capillary bridge
    they feel no force when they approach each other. 
  \item[iii. ] They collide elastically, and upon collision a
    capillary bridge is formed instantaneously. When the particle
    separate, this bridge gives rise to an attractive force which is
    modeled as a potential with a depth \vareps, and a finite range
    \scrit.
  \item[iv. ] The liquid bridge ruptures at the critical rupture
    separation, \scrit, is modeled by removing the potential
    well. When the particles approach the next time they feel no force
    again. In this manner the capillary bridge energy, \vareps, is
    dissipated whenever a capillary bridge is removed (i.e., ruptures)
    after a collision.
\end{itemize} 
Work focusing on individual collisions
\cite{DavisRagerGood2002,AntonyukHeinrichDeenKuipers2009,DonahueHrenyaDavis2010,DonahueHrenyaDavisNakagawaZelinskayaJoseph2010,DonahueBrewerDavisHrenya2012,DonahueDavisKantakHrenya2012,GollwitzerRehbergKruelle2012}
reports a multitude of features for particle collisions involving
capillary interactions that can not fully be captured by this
model. On the other hand, 
the positions of phase boundaries of wet granular fluids appear to be
universal in the sense that they only depend on \vareps\ and \scrit,
and not on other details of the particle interaction \cite{HuangRoeller09}.
For computational convenience all simulations shown in the present
paper therefore adopt a hysteretic square-well potential, i.e., we use
an event-driven algorithm where the potential takes the form of a
square-well with hysteresis as outlined in~i.--iv.

\subsection{Dimensionless units} 

Masses are measured in units of the particle mass, $m$, distances in
units of the particle diameter, $d$, and the time unit is fixed by
measuring forces in terms of $mg$. Non-dimensionalised quantities are
denoted by a hat.
Unless stated otherwise the system size is 
$\hat L \times 2\hat L \times \hat H = 60 \times 120 \times 7.5$, 
and the number of monodisperse particles is
$4.39\times 10^4$, resulting in a filling height of $\hat h=5.0$. Further, the
capillary interaction amounts to $\hatvareps=3/8$ and $\hatscrit=1/16$.

\subsection{Shear flow and arrest} 

A shear flow is induced by applying a space-dependent external force
field
\begin{eqnarray}
  \vec{F}_s (x) 
  &=&  F_s(x) \: \hat{y}
  \nonumber \\[1mm]
  \textrm{ with } \quad 
   F_s(x) 
  &=& \Fsmax \, \cos\frac{ 2\pi x }{L} 
  \label{eq:forcing}
\end{eqnarray}
to the system \cite{Hoover1983,Schulz03,Schulz2006,vollmer09,vollmer2010},
which accelerates particles along the $\hat y$ direction. 
The particles are initially homogeneously distributed within the system with a
Gaussian velocity distribution of mean granular temperature
$T_g/\vareps=40.0$.

For external forces with an amplitude, \Fsmax, slightly larger than the critical forcing,
\Fcrit, the system settles down into a stationary fluid flow whose
local center-of-mass velocity follows the external field.
In \Fig{Snapshot}(b) we show a system with 
a relatively large filling height, $\hat h \simeq 8.8$, and an external force
only $11$\% above \Fcrit. In that situation shear bands form such that the
capillary bridges in the region around $L_x/4$ and $3L_x/4$ are ruptured,
while in the other parts the network of capillary bridges evolves only slowly.
For shear forces smaller than \Fcrit\ the system eventually arrests in
a solid state with a frozen network of nearest neighbors.

\begin{figure} 
\centering
\includegraphics{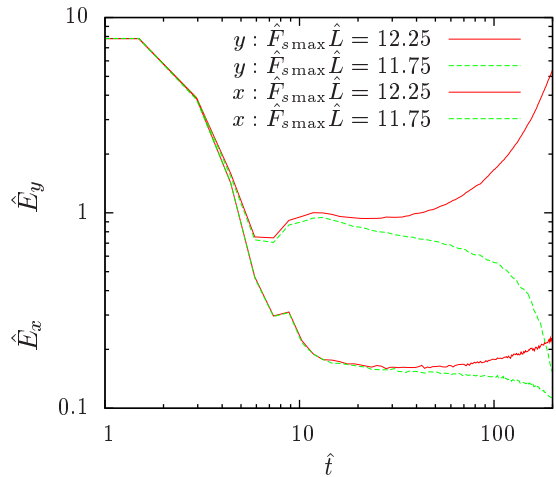}
\caption{Time evolution of the kinetic energy, $E_y$, of motion
  parallel to $\vec F_s$ (upper set of lines), and the
  one in transverse direction, $E_x$ (lower set of lines) for a
  system of size $\hat L = 50$ driven by shear forces with amplitude
  $\hatFsmax\simeq 0.235$ and $0.245$, respectively.}
\label{fig:E_evolution}
\end{figure}

\subsection{Measuring \Fcrit}

The kinetic energy, $E_y$, of motion parallel to the driving shear force will
serve as the order parameter to distinguish the dynamics 
\footnote{In \cite{vollmer2010} we rather used the amplitude of the
  velocity profile in the direction of the external field (i.e., the
  amplitude of the velocity response in reaction to the applied force
  field, \Fsmax, as an order parameter. This has advantages when
  following the hysteresis loop of the response upon slowly decreasing
  and subsequently increasing \Fsmax. However, in the present study,
  where we focus on the arrest of flow, the kinetic energy, $E_y$,
  turned out to be a numerically stable and easier accessible order
  parameter.}.
In \Fig{E_evolution} its time evolution is shown together with the one
of the kinetic energy, $E_x$, of the motion transverse to the external
field. 
When started in a state with high kinetic energy, the fluid first
cools down in a manner closely reminiscent to free cooling
\cite{ulrich09}.  Starting at $\hat t \simeq 10$ it has cooled so far
that the acceleration due to the external field becomes noticeable
over the initial kinetic energy.  Depending on whether the external
field is stronger or weaker than a sharp critical value,
$\hatFcrit\simeq 0.240$, the fluid either gains sufficient energy to
remain in the fluid state forever (cf.~\cite{roeller2009} for the
dynamics in that state), or it settles into the arrested state.
The phase boundary, \Fcrit, is calculated as the mean value between
the neighboring values of shear forces, \hatFsmax, which approach
different states. For the data shown in \Fig{E_evolution} it amounts
to $\hatFcrit \simeq 0.240$.

The phase boundary hence corresponds to the smallest external force
that still leads to sustained shear flow. We demonstrate in the
following that the flow at this threshold corresponds to motion in a
liquid state where the energy injected by the external field is
\emph{exactly} balanced by dissipation due to rupture of capillary
bridges when particles move past each other. For all investigated
systems the flow remains translationally invariant in $y$-direction; in
accordance with the symmetry of the forcing.  Moreover, density of the
liquid can not be distinguished from the random close packing density,
$\phi_{rcp}$, and the height of the layer is spatially uniform.
Consequently, the number of particles, $n(x)$, in a thin rectangular cuboid
of size $2L\times h \times \textrm{d} x$ aligned parallel to the
external field takes a constant value, $N/L$, where $N$ denotes the
number of particles in the system and $L$ the system size transverse
to the flow. This finding is in line with the expectation that a flow
breaking the symmetries of the system or noticeably expanding the bed
would give rise to higher dissipation.

\section{Fluidisation point of sheared wet granular matter}
\label{sec:fluidisation}

In this section we calculate the power, i.e., the energy injected per
unit time into the kinetic energy of the particles due to their motion
in the external force field.  In a steady state this power is balanced
by the energy dissipation rate due to the inelastic particle
collisions. For external forces close to \Fcrit\ the granular
temperature is small. Therefore, effects due to the thermal motion may
be neglected, and the energy dissipation rate is dominated by the
rupture of capillary bridges when particles pass each other in the
flow field.

\subsection{Energy injection rate}

Let $v_y(x)$ denote the flow profile of the flow established when
applying the critical force 
$F_s (x) = \Fsmax\, \cos(2\pi x/L) 
\stackrel{!}{=} \Fcrit \, \cos(2\pi x/L)$.  
The input power that is injected by means of this external force
acting on the particles is given by
\begin{eqnarray}
  \langle P_{\text{forcing}} \rangle 
  &=& \int_0^{L}\textrm{d} x  \; v_y(x)\, F_s(x)\,n(x)
  \nonumber
  \\
  &=& \frac{N \, \Fcrit}{L} \; 
      \int_0^{L}\textrm{d} x  \; v_y(x)\, \cos\frac{2\pi x}{L} \; .
  \label{eq:P_force_general}
\end{eqnarray}
Up to the factor $ N \, \Fcrit$ this amounts to the leading order Fourier
coefficient of the expansion of the velocity profile $v_y(x)$.
Consequently, the injected power takes the from
\begin{equation}
  \langle P_{\text{forcing}} \rangle  
  = C_f \; N \; \Delta v_y \, \Fcrit
  \label{eq:P_force_estimate_harmonic_plug}
\end{equation}
where $\Delta v_y$ is the amplitude of the velocity field. 
Admissible values of $C_f$ lie in the narrow range $0.5 <
C_f < \pi/4$.
After all, the constant $C_f$ takes the value $0.5$ when the velocity
profile is faithfully approximated by its first even harmonic, \ie
$v_y(x) \simeq \Delta v_y\,\cos(2\pi\,x/L)$, and $C_f = \pi/4$ in the
other extreme case of plug flow.

\subsection{Energy dissipation rate}

For each particle the creeping flow enforces a change of
neighbors (in the direction of the flow) with a rate $\dot
\gamma=\textrm{d} v_y/\textrm{d} x$.  Such a displacement goes along
with an energy dissipation $\nu\vareps$ due to rupturing $\nu$
capillary bridges. The total power dissipated upon rupturing capillary
bridges is thus given by
\begin{equation}
  \langle P_{\text{bridge}} \rangle 
  = \int_0^L \textrm{d} x 
  \; n(x) \, \left| \frac{\textrm{d} v_y}{\textrm{d} x} \right|\,\nu\,\vareps~.
  \label{eq:P_diss_general}
\end{equation}
For every $L$-periodic function $v_y(x)$ with a single maximum this
integral yields
\begin{equation}
  \langle P_{\text{bridge}} \rangle 
  =
   \frac{4\,N\,\Delta v_y}{L} \; \nu\,\vareps,
  \label{eq:P_diss_estimate}
\end{equation}
for spatially uniform, $n(x) \equiv N/L$.

For the present system rearrangements are achieved by a slight
vertical expansion of the particle bed.
Assuming that there is no height preference for the rearrangements, and that
there are on average $h/2d$ particles in the column on top of the pair
under consideration, a potential energy of
\begin{equation*}
   U = C_U\; \frac{h}{2d} \; mg \: \Delta h 
\end{equation*}
is associated to the expansion. Here, $C_U$ accounts for the number of columns
to be lifted, and $\Delta h$ to the expansion in height.
Due to the frequent collisions in the dense bed the potential energy
$U$ is immediately dissipated into thermal degrees of freedom of the
granular fluid, and therefore it is not just a one-off
investment. Rather, work has to be done against gravity each time
particles move past each other. Multiplying this energy with the
frequency of particle passages, $4N \Delta v_y/L$ [as given by
\Eq{P_diss_estimate}], therefore provides a second contribution to the
energy dissipation,
\begin{equation}
  \langle P_{\text{grav}} \rangle  
  = \frac{4\,N\,\Delta v_y}{L} \; U
  = \frac{4\,N\,\Delta v_y}{L} \; C_U \; \frac{\Delta h}{2 d} \; m g h \, .
  \label{eq:P_force_grav}
\end{equation}

\subsection{Predicting the critical force \hatFcrit}

We now observe that in a steady state the total dissipation rate due to
rupturing bridges and relaxing $U$ into thermal degrees of freedom has
to balance the input power. We hence obtain
\begin{subequations}
\begin{eqnarray}
  \langle P_{\text{forcing}} \rangle 
  &=& \langle P_{\text{bridge}} \rangle 
  + \langle P_{\text{grav}} \rangle  
\nonumber \\[1mm]
\Leftrightarrow \qquad\qquad
   \Fcrit L
&=& a \; \vareps
+ b \: mgh 
\nonumber \\[1mm]
\Leftrightarrow \qquad\qquad
\hatFcrit \hat{L} 
&=& a \, \hatvareps + b \hat h
\, .
\\[1mm]
\textrm{with } \quad
a &=& \frac{4 \nu}{C_f}
\label{eq:def-a}
\\[1mm]
b &=& \frac{2 C_U \Delta\hat h}{C_f} \, .
\label{eq:def-b}
\end{eqnarray}%
\label{eq:SolidFluidMCM}%
\end{subequations}
We hence predict that \hatFcrit\ is inversely proportional to the
system size $\hat L$, and that $\hatFcrit \hat L$ is a \emph{linear}
function of \hatvareps\ and the filling height $\hat h$. In the
following section we compare these predictions to the results of the
molecular-dynamics simulations.

\begin{figure*} 
\centering
  \includegraphics{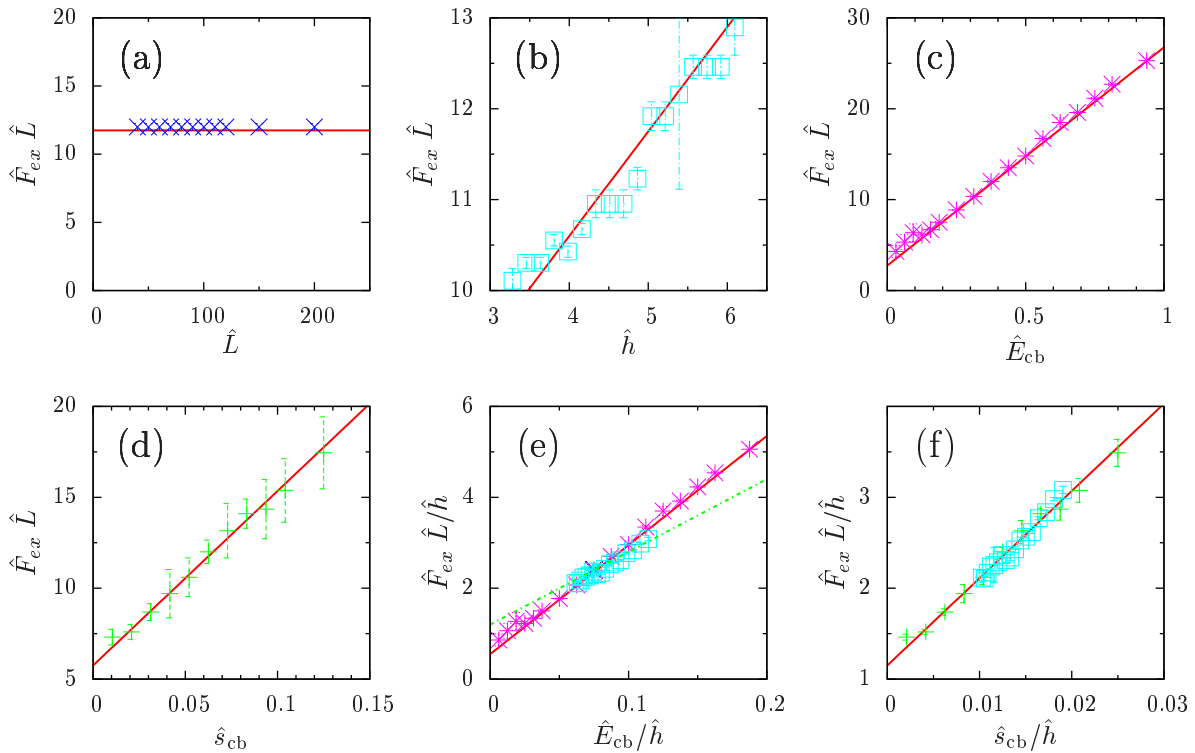}
  \caption{The parameter dependence of the critical driving forces, $\Fsmax$, 
    separating regions that lead to either solid or fluidized
    behavior. The symbols indicate the phase boundary between the two
    phases, and straight lines are the theoretical predictions 
    described in the main text.
    For most of the data points the numerical error in the phase boundary
    is smaller than the size of the symbol.  
    Panels (a)--(d) show the phase boundary between the solid and the
    fluidized state upon varying different parameters: (a) system
    size, $\hat L$, (b) the filling height $\hat h$, (c) dissipated
    energy, $\hatvareps$, and (d) critical rupture
    separation, $\hatscrit$.  In panel (e) $\hatFcrit \hat L/\hat
    h$ is plotted as a function of $\hatvareps/\hat h$ in order to
    demonstrated that the data of panels (b) and (c) are not
    compatible with the prediction, \Eq{SolidFluidMCM}.  Panel (f)
    shows a data collapse of the data of panels (b) and (d), when
    $\hatFcrit \hat L/\hat h$ is plotted as a function
    $\hatscrit/\hat h$ in order to demonstrate that $\hatscrit$ and
    $\hat h$, rather than $\hatvareps$ and $\hat h$ may considered
    independent parameters to predict $\hatFcrit \hat L$.  The method
    for performing the simulations is discussed in the text.}
\label{fig:md_test}
\end{figure*}

\begin{figure}
  \centering
  \includegraphics{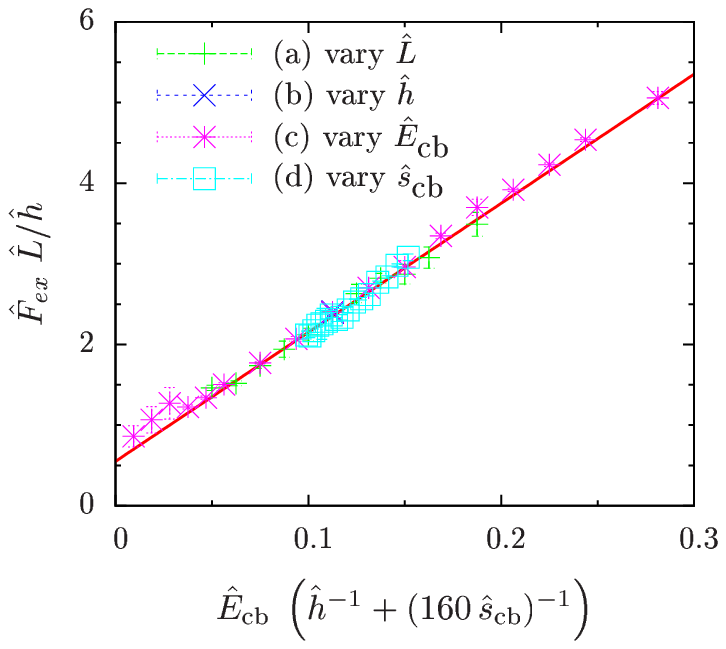}
  \caption{Master plot incorporating all data shown in the
      different panels of \Fig{md_test}.  The dependence of the
      critical force \Fcrit\ on the system size, $\hat{L}$, the
      filling height, $\hat h$, the energy dissipated upon rupturing
      capillary bridges, \hatvareps, and the critical rupture
      separation,\hatscrit, is faithfully described by \Eq{theory}.
      The solid line shows a theoretical prediction, \Eq{theory}
      involving only three free parameters $\tilde a$, $\tilde b$, and
      $\tilde\chi$ whose values can also explicitly be calculated.}
  \label{fig:master}
\end{figure}

\section{Phase Diagrams}
\label{sec:phase-diagram}

In \Fig{md_test}(a--d) we explore the dependence of \hatFcrit\ on the
systems size, $L$, the filling height, $h$, the dissipated energy,
$\vareps$, and the rupture length, \hatscrit. Plotting \hatFcrit\ as a
function of the respective parameters provides sections through the
phase diagram:  there is sustained flow for values of $\hat F_s$
larger than $\hatFcrit$, and flow is arrested for smaller external
forces.

\subsection{Parameter dependence of \hatFcrit}

(a) The variation of the system size $\hat L$ was done whilst keeping the
aspect ratio of the container constant at $L_y/L_x=2$. At the same time the
average particle number density was kept constant at $\phi=0.43$ which means
that the number of particles is changing in order to provide a fixed
filling height, $\hat h \simeq 5$.
As expected from \Eq{SolidFluidMCM} the value $\hatFcrit \hat L$ is
constant.
We find that $\hatFcrit \hat L \simeq 12$ for fixed $\hatvareps = 0.375$
and $\hat h \simeq 5$.

(b) The filling height, $h$, was varied by changing the number of particles
in the system whilst keeping the geometry of the simulation volume and particle
interactions fixed.
The filling height is estimated as the filling height in the solid state when
assuming random close packing.
As predicted by \Eq{SolidFluidMCM} the dependence of $\hatFcrit \hat L$
on $\hat h$ for a fixed $\hatvareps$ is linear. In the simulations we
find
\begin{equation}
\hatFcrit \hat{L}   \simeq   6.0 + 1.2 \; \hat h
\label{eq:FL-h}
\end{equation}
for $\hatvareps = 0.375$.

(c) Whilst varying the capillary bridge energy, \vareps, the rupture
separation, $\hatscrit = 1/16$, filling height, $\hat h \simeq 5$, and the
system size, $\hat L = 60$, were kept constant.
\Fig{md_test}(c) shows the expected linear dependence of $\hatFcrit \hat L$
on $\hatvareps$, 
\begin{equation}
\hatFcrit \hat{L}   \simeq  24 \hatvareps + 2.8 \, .
\label{eq:FL-E}
\end{equation}

(d) The variation of the critical rupture separation, $\scrit$, was done for a
fixed aspect ratio of width vs.~depth of the potential well, i.e.~$\vareps$
was varied together with \scrit\ at a fixed ratio of $R\equiv \vareps/\scrit=6.0$. System
size and filling height were fixed to $\hat L=60$ and $\hat h=5$,
respectively. 
The resulting linear dependence 
\begin{equation}
  \hatFcrit \hat{L}   \simeq  16\, R\, \hatscrit + 6.0
\label{eq:FL-s}
\end{equation}
is shown in \Fig{md_test}(d).

\subsection{Consistency checks}

Merely finding linear dependences of \hatFcrit\ on $\hat h$,
\hatvareps, and \hatscrit\ is not sufficient to show that
\Eq{SolidFluidMCM} faithfully describes the arrest of flow. One also
has to verify that the linear functions are mutually consistent for
all sections through the phase diagram.

(e) Observing that the $y$-intercept in \Eq{FL-h} may be written as
$6.0 \simeq 16\,\hatvareps$ for $\hatvareps = 0.375$, and that the one
in \Eq{FL-E} is $2.8 \simeq 0.56\, \hat h$ for $\hat h = 5$, one
realizes, that the linear dependencies, \Eqs{FL-h} and \eq{FL-E}, can
not be compatible with \Eq{SolidFluidMCM} at the same time.  After
all, $0.56$ disagrees with $1.2$, and $24$ disagrees with $16$. In
\Fig{md_test}(e) this is demonstrated by plotting $\hatFcrit \hat{L}$
versus $\hatvareps / \hat h$: the data shown in \Fig{md_test}(b) and
\Fig{md_test}(c) lie on different straight lines.

(f) On the other hand the data shown in panels (d) and (b) are
compatible. \Eq{FL-s} is commensurate with \Eq{FL-h} since the latter
has a $y$-intercept of $6.0 = 16 \hatvareps = 16 R \hatscrit$ with
$R=6.0$ and $\hatscrit=1/16$, and since the $y$-intercept of \Eq{FL-s}
is $6.0 = 1.2 \hat h$ with $\hat h = 5.0$.

These findings suggest that $\hatvareps$ and $\hat h$ are not
independent variables --- as put forward by \Eq{SolidFluidMCM} when
assuming that $a$ and $b$ take constant values. 
In the following we show that a consistent description of the
numerical data can be achieved, however, by assuming that $a$ has a
weak, linear dependence on $\hat h$. It reflects that the number of
capillary bridges, $\nu$, ruptured in an exchange event depends on the
filling height $\hat h$.

\section{Data Collapse}
\label{sec:discussion}

Equation \eq{FL-s} can only be consistent with \Eq{FL-E} if
its $y$-intercept comprises a contribution proportional to $R$, and if
the slopes are adjusted accordingly, \ie by decomposing the
$y$-intercept as
$6.0 = 1.2 \, \hat h \simeq ( 0.55 + R/10 ) \, \hat h$ with $\hat h = 5$ and $R=6.0$. 
Observing that $R\equiv \vareps/\scrit$ this provides an improved prediction
\begin{subequations}
\begin{eqnarray}
  \hatFcrit  \hat L
  & \simeq &
  \left( 1 +  \tilde\chi \; \frac{\hat{h}}{\hatscrit}\right) 
  \tilde{a} \, \hatvareps \; 
  + \tilde{b}  \hat h
\, ,\\
\textrm{with \ } 
\tilde{a} &\simeq& 16 \, ,
\label{eq:atilde}
\\ 
\tilde{b} &\simeq& 0.55
\label{eq:btilde}
\\
\tilde{\chi} &\simeq& 1/160
\label{eq:chi}
\end{eqnarray}%
\label{eq:theory}%
\end{subequations}
which takes into account the increase of the number of bridge ruptures
upon increasing $h/\scrit$ due to the possibility to rupture bridges
in the vertically displaced column of particles on top of the site
where a swapping event occurs. As demonstrated in \Fig{master} this
prediction is in excellent agreement with all data.
As a final step of data analysis we interpret the values of the fitting
parameters $\tilde a$, $\tilde b$ and $\tilde\chi$ entering \Eq{theory}.

\subsection{Determine $\tilde a$}

In view of \Eq{def-a}
${a}$ is related to the number, $\nu$, of capillary
bridge ruptures in every particle exchange 
\begin{equation}
  {a} = \frac{4 \, \nu}{C_f} \simeq 8 \nu
\end{equation}
where we used the estimate $C_f \simeq 0.5$, as argued upon
introducing this constant in
\Eq{P_force_estimate_harmonic_plug}. Moreover, in order to take into
account the correction term for the height dependence introduced in
\Eq{theory} another factor $1+\hat h/160 \hatscrit$ must be accounted
for. 
For the standard choice of parameters $\hat h=5$ and $\hatscrit =
1/16$ we hence find 
\begin{equation}
  \tilde{a} 
= \frac{4 \; \nu}{C_f} \; \left( 1+ \frac{\hat h}{160 \hatscrit} \right)^{-1}  
\simeq \frac{16}{3} \; \nu
\end{equation}
Finally, in a granular bed that is expanding to allow particles to
pass each other the number $\nu$ of rupture events must be larger than~$1$, 
but still small. For $\nu = 3$ we hence recover the value 
$\tilde a = 16$ reported in \Eq{atilde}.

\subsection{Determine $\tilde b$}

In order to see that the value for $b \simeq 0.55$ is
reasonable, too, we observe that subsequent rows of spheres in a close-packed structure
are separated by a height distance $\hat h_{\textrm{cp}}=\sqrt{2/3}\simeq
0.816$ and that the saddle that has to be passed to roll from one minimum to a
nearby minimum is at height $\hat h_{\textrm{saddle}}=\sqrt{3}/2 \simeq
0.866$. The minimum expansion in height to move over the potential landscape
set up by the layers below is therefore of the order of $\Delta\hat h\simeq
\hat h_{\textrm{saddle}} - \hat h_{\textrm{cp}} \simeq 0.05$, and 
potential wells in a disordered, only slightly expanded random packing, will
still be of the same order of magnitude. Based on this estimate, on $C_f\simeq
0.5$, and on \Eq{def-b} the number of columns, $C_U$, lifted a swapping event
amounts to
\begin{equation}
  C_U 
  \simeq 0.55 \frac{C_f}{2 \Delta\hat h} 
  \simeq 2.8 \, .
\end{equation}
Lifting a small number of columns lends plausibility to $b\simeq 0.55$.

\subsection{Determine $\tilde\chi$}

In order to gain insight into the order of magnitude of $\tilde\chi$
we note that the lifted column needs to rupture bonds all along its
walls.  Hence, $\nu$ is not merely dependent on its cross-section, as
implied by assuming $\nu$ to be independent of $\hat{h}$. The number
of bonds that are broken is then expected to scale linearly with
column height and inversely proportional to \hatscrit. After all,
bonds are allowed to stretch to a finite length \hatscrit, and the
larger \hatscrit\ the lower the likelihood that the dilation requires
a given bond (along the vertical walls of the pile of particles
displaced vertically) to be broken.  This leads to an additional
number $\tilde\chi \nu \hat{h}/\hatscrit$ of bonds broken per column.
As we saw above, it increases the number of bond ruptures by about
$50$\% which seems reasonable in a system with a packing height of
only five layers and a rupture separation, $\hatscrit = 1/16 = 0.065$
that is larger than the height displacement, $\Delta \tilde h=0.05$,
of the column. After all, in such a situation only pre-stretched bonds
are likely to rupture.
Keeping this in mind, the small value of $\tilde\chi$ may be
understood as a result of incorporating a factor of $\Delta \tilde h$ and a
probability of about $1/4$ to encounter a pre-stretched bond in a
layer of a column that is vertically displaced. In this interpretation
the average number of layers in the column amounts to $\hat h/2$.

\section{Summary and Outlook}
\label{sec:conclusion}

The present study substantiates the finding that the transition from a
fluidized to an arrested state in wet granular matter arises when the
dissipation rate due to the rupture of capillary bridges in the shear
flow can no longer be balanced by the power injection from the
external field.
Earlier work \cite{vollmer2010} showed that this approach provides a
comprehensive understanding of the transition in the setting of a
two-dimensional flow of bidisperse disks where the density is fixed by
a confining box.
The present work addressed the flow of a bed of monodisperse spheres
that are confined in the vertical direction by a gravitational field.
Closely above the transition the flow is reminiscent to a
slow plastic flow of the bed in the direction of the applied
field. The granular bed expands only minimally in vertical
direction. For the small flow velocities gravity still keeps density,
filling height, and pressure to values observed in an arrested
packing.
Also for this setting, which fundamentally differs 
from systems where the particles are confined by walls, 
the power balance provides an excellent prediction, \Eq{theory}, of the
critical force where flow ceases. This is demonstrated by a master plot,
\Fig{master}, showing an excellent data collapse of numerical data
obtained by varying four different characteristics of the system: the
system size, the filling height, as well as the strength and critical
rupture separation of capillary bridges.
This data collapse suggests 
\begin{enumerate}
\item[i. ] When the granular flow is confined by gravity, typically
  only $2$--$3$ capillary bridges are ruptured upon swapping
  neighboring particles moving with slightly different speed in the
  direction of the external forcing. This is a striking difference to
  wall-bounded flows \cite{vollmer2010} where this number
  diverges when the density approaches close packing.
\item[ii. ] In a gravity-confined granular bed the effortless passing
  of the particles is facilitated
  by a slight expansion of the granular bed where $2$--$3$ columns of
  particles are lifted by a small amount to let the particles pass
  between neighboring potential wells arising from the corrugations
  formed by the layer below. The associated potential energy is also
  dissipated. 
\item[iii. ] With a small probability additional capillary bridges are
  broken due to the slight expansions in vertical direction.
\end{enumerate}
The prediction of the flow-threshold involves only three constants
characterizing the processes~i.--iii. The values of these constants
have been determined to a very good accuracy in
\Sec{discussion}. 

The most remarkable finding of our study is that the critical
force \Fcrit\ does \emph{not} depend on the specific form of the flow
profile. It can be calculated without specifying the hydrodynamic
equations of the flow and determining their solution.
We therefore conclude that 
the approach of balancing the energy input rate (due to
the external force inducing the flow) 
and the dissipation rate (due the
rupture of capillary bridges, when particles move past each other)
provides a powerful framework to study the arrest of flow in wet granular
materials where dissipation is dominated by capillary bridge ruptures.
This approach provides a universal framework to predict the threshold
for the arrest of flow, and it can be applied without need to
determine flow profiles.
Forthcoming work will explore this intriguing possibility also for
flows in other geometries and due to other forcing.

\begin{acknowledgments}

  We are indebted to Martin Brinkmann, Karen Daniels and
  Matthias Schr\"oter for stimulating discussion, and to our two 
  referees for very helpful comments to improve the presentation of
  our results. 

\end{acknowledgments}


\begin{thebibliography}{49}%
\makeatletter
\providecommand \@ifxundefined [1]{%
 \@ifx{#1\undefined}
}%
\providecommand \@ifnum [1]{%
 \ifnum #1\expandafter \@firstoftwo
 \else \expandafter \@secondoftwo
 \fi
}%
\providecommand \@ifx [1]{%
 \ifx #1\expandafter \@firstoftwo
 \else \expandafter \@secondoftwo
 \fi
}%
\providecommand \natexlab [1]{#1}%
\providecommand \enquote  [1]{``#1''}%
\providecommand \bibnamefont  [1]{#1}%
\providecommand \bibfnamefont [1]{#1}%
\providecommand \citenamefont [1]{#1}%
\providecommand \href@noop [0]{\@secondoftwo}%
\providecommand \href [0]{\begingroup \@sanitize@url \@href}%
\providecommand \@href[1]{\@@startlink{#1}\@@href}%
\providecommand \@@href[1]{\endgroup#1\@@endlink}%
\providecommand \@sanitize@url [0]{\catcode `\\12\catcode `\$12\catcode
  `\&12\catcode `\#12\catcode `\^12\catcode `\_12\catcode `\%12\relax}%
\providecommand \@@startlink[1]{}%
\providecommand \@@endlink[0]{}%
\providecommand \url  [0]{\begingroup\@sanitize@url \@url }%
\providecommand \@url [1]{\endgroup\@href {#1}{\urlprefix }}%
\providecommand \urlprefix  [0]{URL }%
\providecommand \Eprint [0]{\href }%
\providecommand \doibase [0]{http://dx.doi.org/}%
\providecommand \selectlanguage [0]{\@gobble}%
\providecommand \bibinfo  [0]{\@secondoftwo}%
\providecommand \bibfield  [0]{\@secondoftwo}%
\providecommand \translation [1]{[#1]}%
\providecommand \BibitemOpen [0]{}%
\providecommand \bibitemStop [0]{}%
\providecommand \bibitemNoStop [0]{.\EOS\space}%
\providecommand \EOS [0]{\spacefactor3000\relax}%
\providecommand \BibitemShut  [1]{\csname bibitem#1\endcsname}%
\let\auto@bib@innerbib\@empty
\bibitem [{\citenamefont {{GDR MiDi}}(2004)}]{MiDi2004}%
  \BibitemOpen
  \bibfield  {author} {\bibinfo {author} {\bibnamefont {{GDR MiDi}}},\ }\href
  {\doibase 10.1140/epje/i2003-10153-0} {\bibfield  {journal} {\bibinfo
  {journal} {Eur. Phys. J. E}\ }\textbf {\bibinfo {volume} {14}},\ \bibinfo
  {pages} {341} (\bibinfo {year} {2004})}\BibitemShut {NoStop}%
\bibitem [{\citenamefont {Jaeger}\ \emph {et~al.}(1996)\citenamefont {Jaeger},
  \citenamefont {Nagel},\ and\ \citenamefont
  {Behringer}}]{JaegerNagelBehringer1996}%
  \BibitemOpen
  \bibfield  {author} {\bibinfo {author} {\bibfnamefont {H.~M.}\ \bibnamefont
  {Jaeger}}, \bibinfo {author} {\bibfnamefont {S.~R.}\ \bibnamefont {Nagel}}, \
  and\ \bibinfo {author} {\bibfnamefont {R.~P.}\ \bibnamefont {Behringer}},\
  }\href {\doibase 10.1103/RevModPhys.68.1259} {\bibfield  {journal} {\bibinfo
  {journal} {Rev. Mod. Phys.}\ }\textbf {\bibinfo {volume} {68}},\ \bibinfo
  {pages} {1259} (\bibinfo {year} {1996})}\BibitemShut {NoStop}%
\bibitem [{\citenamefont {Kadanoff}(1999)}]{kadanoff99}%
  \BibitemOpen
  \bibfield  {author} {\bibinfo {author} {\bibfnamefont {L.~P.}\ \bibnamefont
  {Kadanoff}},\ }\href {\doibase 10.1103/RevModPhys.71.435} {\bibfield
  {journal} {\bibinfo  {journal} {Rev. Mod. Phys.}\ }\textbf {\bibinfo {volume}
  {71}},\ \bibinfo {pages} {435} (\bibinfo {year} {1999})}\BibitemShut
  {NoStop}%
\bibitem [{\citenamefont {Silbert}\ \emph {et~al.}(2001)\citenamefont
  {Silbert}, \citenamefont {Erta\c{s}}, \citenamefont {Grest}, \citenamefont
  {Halsey}, \citenamefont {Levine},\ and\ \citenamefont
  {Plimpton}}]{SilbertErtas01}%
  \BibitemOpen
  \bibfield  {author} {\bibinfo {author} {\bibfnamefont {L.~E.}\ \bibnamefont
  {Silbert}}, \bibinfo {author} {\bibfnamefont {D.}~\bibnamefont {Erta\c{s}}},
  \bibinfo {author} {\bibfnamefont {G.~S.}\ \bibnamefont {Grest}}, \bibinfo
  {author} {\bibfnamefont {T.~C.}\ \bibnamefont {Halsey}}, \bibinfo {author}
  {\bibfnamefont {D.}~\bibnamefont {Levine}}, \ and\ \bibinfo {author}
  {\bibfnamefont {S.~J.}\ \bibnamefont {Plimpton}},\ }\href {\doibase
  10.1103/PhysRevE.64.051302} {\bibfield  {journal} {\bibinfo  {journal} {Phys.
  Rev. E}\ }\textbf {\bibinfo {volume} {64}},\ \bibinfo {pages} {051302}
  (\bibinfo {year} {2001})}\BibitemShut {NoStop}%
\bibitem [{\citenamefont {Aranson}\ and\ \citenamefont
  {Tsimring}(2006)}]{Aranson2006}%
  \BibitemOpen
  \bibfield  {author} {\bibinfo {author} {\bibfnamefont {I.~S.}\ \bibnamefont
  {Aranson}}\ and\ \bibinfo {author} {\bibfnamefont {L.~S.}\ \bibnamefont
  {Tsimring}},\ }\href {\doibase 10.1103/RevModPhys.78.641} {\bibfield
  {journal} {\bibinfo  {journal} {Reviews of Modern Physics}\ }\textbf
  {\bibinfo {volume} {78}},\ \bibinfo {eid} {641} (\bibinfo {year}
  {2006})}\BibitemShut {NoStop}%
\bibitem [{\citenamefont {Jop}\ \emph {et~al.}(2006)\citenamefont {Jop},
  \citenamefont {Forterre},\ and\ \citenamefont
  {Pouliquen}}]{JopForterrePouliquen2006}%
  \BibitemOpen
  \bibfield  {author} {\bibinfo {author} {\bibfnamefont {P.}~\bibnamefont
  {Jop}}, \bibinfo {author} {\bibfnamefont {Y.}~\bibnamefont {Forterre}}, \
  and\ \bibinfo {author} {\bibfnamefont {O.}~\bibnamefont {Pouliquen}},\ }\href
  {\doibase 10.1038/nature04801} {\bibfield  {journal} {\bibinfo  {journal}
  {Nature}\ }\textbf {\bibinfo {volume} {441}},\ \bibinfo {pages} {727}
  (\bibinfo {year} {2006})}\BibitemShut {NoStop}%
\bibitem [{\citenamefont {Borzsonyi}\ and\ \citenamefont
  {Ecke}(2007)}]{BorzsonyiEcke2007}%
  \BibitemOpen
  \bibfield  {author} {\bibinfo {author} {\bibfnamefont {T.}~\bibnamefont
  {Borzsonyi}}\ and\ \bibinfo {author} {\bibfnamefont {R.~E.}\ \bibnamefont
  {Ecke}},\ }\href {\doibase 10.1103/PhysRevE.76.031301} {\bibfield  {journal}
  {\bibinfo  {journal} {Phys. Rev. E}\ }\textbf {\bibinfo {volume} {76}},\
  \bibinfo {eid} {031301} (\bibinfo {year} {2007})}\BibitemShut {NoStop}%
\bibitem [{\citenamefont {Forterre}\ and\ \citenamefont
  {Pouliquen}(2008)}]{ForterrePouliquen2008}%
  \BibitemOpen
  \bibfield  {author} {\bibinfo {author} {\bibfnamefont {Y.}~\bibnamefont
  {Forterre}}\ and\ \bibinfo {author} {\bibfnamefont {O.}~\bibnamefont
  {Pouliquen}},\ }\href {\doibase 10.1146/annurev.fluid.40.111406.102142}
  {\bibfield  {journal} {\bibinfo  {journal} {Annual Review of Fluid
  Mechanics}\ }\textbf {\bibinfo {volume} {40}},\ \bibinfo {pages} {1}
  (\bibinfo {year} {2008})}\BibitemShut {NoStop}%
\bibitem [{\citenamefont {Luding}(2009)}]{Luding2009}%
  \BibitemOpen
  \bibfield  {author} {\bibinfo {author} {\bibfnamefont {S.}~\bibnamefont
  {Luding}},\ }\href@noop {} {\bibfield  {journal} {\bibinfo  {journal}
  {Nonlinearity}\ }\textbf {\bibinfo {volume} {22}},\ \bibinfo {pages} {R101}
  (\bibinfo {year} {2009})}\BibitemShut {NoStop}%
\bibitem [{\citenamefont {Utter}\ and\ \citenamefont
  {Behringer}(2008)}]{UtterBehringer2008}%
  \BibitemOpen
  \bibfield  {author} {\bibinfo {author} {\bibfnamefont {B.}~\bibnamefont
  {Utter}}\ and\ \bibinfo {author} {\bibfnamefont {R.~P.}\ \bibnamefont
  {Behringer}},\ }\href {\doibase 10.1103/PhysRevLett.100.208302} {\bibfield
  {journal} {\bibinfo  {journal} {Phys. Rev. Lett.}\ }\textbf {\bibinfo
  {volume} {100}},\ \bibinfo {pages} {208302} (\bibinfo {year}
  {2008})}\BibitemShut {NoStop}%
\bibitem [{\citenamefont {Berardi}\ \emph {et~al.}(2010)\citenamefont
  {Berardi}, \citenamefont {Barros}, \citenamefont {Douglas},\ and\
  \citenamefont {Losert}}]{BerardiBarrosDouglasLosert2010}%
  \BibitemOpen
  \bibfield  {author} {\bibinfo {author} {\bibfnamefont {C.~R.}\ \bibnamefont
  {Berardi}}, \bibinfo {author} {\bibfnamefont {K.}~\bibnamefont {Barros}},
  \bibinfo {author} {\bibfnamefont {J.~F.}\ \bibnamefont {Douglas}}, \ and\
  \bibinfo {author} {\bibfnamefont {W.}~\bibnamefont {Losert}},\ }\href
  {\doibase 10.1103/PhysRevE.81.041301} {\bibfield  {journal} {\bibinfo
  {journal} {Phys. Rev. E}\ }\textbf {\bibinfo {volume} {81}},\ \bibinfo
  {pages} {041301} (\bibinfo {year} {2010})}\BibitemShut {NoStop}%
\bibitem [{\citenamefont {van Hecke}(2010)}]{Hecke2010}%
  \BibitemOpen
  \bibfield  {author} {\bibinfo {author} {\bibfnamefont {M.}~\bibnamefont {van
  Hecke}},\ }\href {\doibase 10.1088/0953-8984/22/3/033101} {\bibfield
  {journal} {\bibinfo  {journal} {Journal of Physics: Condensed Matter}\
  }\textbf {\bibinfo {volume} {22}},\ \bibinfo {pages} {033101} (\bibinfo
  {year} {2010})}\BibitemShut {NoStop}%
\bibitem [{\citenamefont {Tordesillas}\ \emph {et~al.}(2011)\citenamefont
  {Tordesillas}, \citenamefont {Lin}, \citenamefont {Zhang}, \citenamefont
  {Behringer},\ and\ \citenamefont
  {Shi}}]{TordesillasLinZhangBehringerShi2011}%
  \BibitemOpen
  \bibfield  {author} {\bibinfo {author} {\bibfnamefont {A.}~\bibnamefont
  {Tordesillas}}, \bibinfo {author} {\bibfnamefont {Q.}~\bibnamefont {Lin}},
  \bibinfo {author} {\bibfnamefont {J.}~\bibnamefont {Zhang}}, \bibinfo
  {author} {\bibfnamefont {R.}~\bibnamefont {Behringer}}, \ and\ \bibinfo
  {author} {\bibfnamefont {J.}~\bibnamefont {Shi}},\ }\href {\doibase
  10.1016/j.jmps.2010.10.007} {\bibfield  {journal} {\bibinfo  {journal}
  {Journal of the Mechanics and Physics of Solids}\ }\textbf {\bibinfo {volume}
  {59}},\ \bibinfo {pages} {265 } (\bibinfo {year} {2011})}\BibitemShut
  {NoStop}%
\bibitem [{\citenamefont {Lois}\ \emph {et~al.}(2009)\citenamefont {Lois},
  \citenamefont {Zhang}, \citenamefont {Majmudar}, \citenamefont {Henkes},
  \citenamefont {Chakraborty}, \citenamefont {O'Hern},\ and\ \citenamefont
  {Behringer}}]{LoisZhangMajmudarHenkesChakrabortyOHernBehringer2009}%
  \BibitemOpen
  \bibfield  {author} {\bibinfo {author} {\bibfnamefont {G.}~\bibnamefont
  {Lois}}, \bibinfo {author} {\bibfnamefont {J.}~\bibnamefont {Zhang}},
  \bibinfo {author} {\bibfnamefont {T.~S.}\ \bibnamefont {Majmudar}}, \bibinfo
  {author} {\bibfnamefont {S.}~\bibnamefont {Henkes}}, \bibinfo {author}
  {\bibfnamefont {B.}~\bibnamefont {Chakraborty}}, \bibinfo {author}
  {\bibfnamefont {C.~S.}\ \bibnamefont {O'Hern}}, \ and\ \bibinfo {author}
  {\bibfnamefont {R.~P.}\ \bibnamefont {Behringer}},\ }\href {\doibase
  10.1103/PhysRevE.80.060303} {\bibfield  {journal} {\bibinfo  {journal} {Phys.
  Rev. E}\ }\textbf {\bibinfo {volume} {80}},\ \bibinfo {pages} {060303}
  (\bibinfo {year} {2009})}\BibitemShut {NoStop}%
\bibitem [{\citenamefont {Trappe}\ \emph {et~al.}(2001)\citenamefont {Trappe},
  \citenamefont {Prasad}, \citenamefont {Cipelletti}, \citenamefont {Segre},\
  and\ \citenamefont {Weitz}}]{TrappePrasadCipellettiSegreWeitz2001}%
  \BibitemOpen
  \bibfield  {author} {\bibinfo {author} {\bibfnamefont {V.}~\bibnamefont
  {Trappe}}, \bibinfo {author} {\bibfnamefont {V.}~\bibnamefont {Prasad}},
  \bibinfo {author} {\bibfnamefont {L.}~\bibnamefont {Cipelletti}}, \bibinfo
  {author} {\bibfnamefont {P.}~\bibnamefont {Segre}}, \ and\ \bibinfo {author}
  {\bibfnamefont {D.~A.}\ \bibnamefont {Weitz}},\ }\href@noop {} {\bibfield
  {journal} {\bibinfo  {journal} {Nature}\ }\textbf {\bibinfo {volume} {411}},\
  \bibinfo {pages} {772 } (\bibinfo {year} {2001})}\BibitemShut {NoStop}%
\bibitem [{\citenamefont {Rognon}\ \emph {et~al.}(2006)\citenamefont {Rognon},
  \citenamefont {Roux}, \citenamefont {Wolf}, \citenamefont {Naa\"im},\ and\
  \citenamefont {Chevoir}}]{RognonRouxWolfNaaimChevoir2006}%
  \BibitemOpen
  \bibfield  {author} {\bibinfo {author} {\bibfnamefont {P.~G.}\ \bibnamefont
  {Rognon}}, \bibinfo {author} {\bibfnamefont {J.-N.}\ \bibnamefont {Roux}},
  \bibinfo {author} {\bibfnamefont {D.}~\bibnamefont {Wolf}}, \bibinfo {author}
  {\bibfnamefont {M.}~\bibnamefont {Naa\"im}}, \ and\ \bibinfo {author}
  {\bibfnamefont {F.}~\bibnamefont {Chevoir}},\ }\href {\doibase
  10.1209/epl/i2005-10578-y} {\bibfield  {journal} {\bibinfo  {journal} {EPL}\
  }\textbf {\bibinfo {volume} {74}},\ \bibinfo {pages} {644} (\bibinfo {year}
  {2006})}\BibitemShut {NoStop}%
\bibitem [{\citenamefont {Rognon}\ \emph {et~al.}(2008)\citenamefont {Rognon},
  \citenamefont {Roux}, \citenamefont {Naa\"im},\ and\ \citenamefont
  {Chevoir}}]{RognonRouxNaaimChevoir2008}%
  \BibitemOpen
  \bibfield  {author} {\bibinfo {author} {\bibfnamefont {P.~G.}\ \bibnamefont
  {Rognon}}, \bibinfo {author} {\bibfnamefont {J.-N.}\ \bibnamefont {Roux}},
  \bibinfo {author} {\bibfnamefont {M.}~\bibnamefont {Naa\"im}}, \ and\
  \bibinfo {author} {\bibfnamefont {F.}~\bibnamefont {Chevoir}},\ }\href
  {\doibase 10.1017/S0022112007009329} {\bibfield  {journal} {\bibinfo
  {journal} {Journal of Fluid Mechanics}\ }\textbf {\bibinfo {volume} {596}},\
  \bibinfo {pages} {21} (\bibinfo {year} {2008})}\BibitemShut {NoStop}%
\bibitem [{\citenamefont {Lois}\ \emph {et~al.}(2007)\citenamefont {Lois},
  \citenamefont {Blawzdziewicz},\ and\ \citenamefont
  {O'Hern}}]{LoisBlawzdziewiczOHernCorey2007}%
  \BibitemOpen
  \bibfield  {author} {\bibinfo {author} {\bibfnamefont {G.}~\bibnamefont
  {Lois}}, \bibinfo {author} {\bibfnamefont {J.}~\bibnamefont {Blawzdziewicz}},
  \ and\ \bibinfo {author} {\bibfnamefont {C.~S.}\ \bibnamefont {O'Hern}},\
  }\href {\doibase 10.1002/pamm.200700510} {\bibfield  {journal} {\bibinfo
  {journal} {PAMM}\ }\textbf {\bibinfo {volume} {7}},\ \bibinfo {pages}
  {1090605} (\bibinfo {year} {2007})}\BibitemShut {NoStop}%
\bibitem [{\citenamefont {Lois}\ \emph {et~al.}(2008)\citenamefont {Lois},
  \citenamefont {Blawzdziewicz},\ and\ \citenamefont
  {O'Hern}}]{LoisBlawzdziewiczOHernCorey2008}%
  \BibitemOpen
  \bibfield  {author} {\bibinfo {author} {\bibfnamefont {G.}~\bibnamefont
  {Lois}}, \bibinfo {author} {\bibfnamefont {J.}~\bibnamefont {Blawzdziewicz}},
  \ and\ \bibinfo {author} {\bibfnamefont {C.~S.}\ \bibnamefont {O'Hern}},\
  }\href {\doibase 10.1103/PhysRevLett.100.028001} {\bibfield  {journal}
  {\bibinfo  {journal} {Phys. Rev. Lett.}\ }\textbf {\bibinfo {volume} {100}},\
  \bibinfo {pages} {028001} (\bibinfo {year} {2008})}\BibitemShut {NoStop}%
\bibitem [{\citenamefont {Pitois}\ \emph {et~al.}(2000)\citenamefont {Pitois},
  \citenamefont {Moucheront},\ and\ \citenamefont
  {Chateau}}]{PitoisMoucherontChateau2000}%
  \BibitemOpen
  \bibfield  {author} {\bibinfo {author} {\bibfnamefont {O.}~\bibnamefont
  {Pitois}}, \bibinfo {author} {\bibfnamefont {P.}~\bibnamefont {Moucheront}},
  \ and\ \bibinfo {author} {\bibfnamefont {X.}~\bibnamefont {Chateau}},\ }\href
  {\doibase 10.1006/jcis.2000.7096} {\bibfield  {journal} {\bibinfo  {journal}
  {Journal of Colloid and Interface Science}\ }\textbf {\bibinfo {volume}
  {231}},\ \bibinfo {pages} {26} (\bibinfo {year} {2000})}\BibitemShut
  {NoStop}%
\bibitem [{\citenamefont {Kantak}\ \emph {et~al.}(2009)\citenamefont {Kantak},
  \citenamefont {Hrenya},\ and\ \citenamefont {Davis}}]{KantakHrenyaDavis2009}%
  \BibitemOpen
  \bibfield  {author} {\bibinfo {author} {\bibfnamefont {A.~A.}\ \bibnamefont
  {Kantak}}, \bibinfo {author} {\bibfnamefont {C.~M.}\ \bibnamefont {Hrenya}},
  \ and\ \bibinfo {author} {\bibfnamefont {R.~H.}\ \bibnamefont {Davis}},\
  }\href {\doibase 10.1063/1.3070830} {\bibfield  {journal} {\bibinfo
  {journal} {Physics of Fluids}\ }\textbf {\bibinfo {volume} {21}},\ \bibinfo
  {pages} {023301} (\bibinfo {year} {2009})}\BibitemShut {NoStop}%
\bibitem [{\citenamefont {Donahue}\ \emph
  {et~al.}(2010{\natexlab{a}})\citenamefont {Donahue}, \citenamefont {Hrenya},\
  and\ \citenamefont {Davis}}]{DonahueHrenyaDavis2010}%
  \BibitemOpen
  \bibfield  {author} {\bibinfo {author} {\bibfnamefont {C.}~\bibnamefont
  {Donahue}}, \bibinfo {author} {\bibfnamefont {C.}~\bibnamefont {Hrenya}}, \
  and\ \bibinfo {author} {\bibfnamefont {R.}~\bibnamefont {Davis}},\ }\href
  {\doibase 10.1103/PhysRevLett.105.034501} {\bibfield  {journal} {\bibinfo
  {journal} {Phys. Rev. Lett.}\ }\textbf {\bibinfo {volume} {105}},\ \bibinfo
  {pages} {034501} (\bibinfo {year} {2010}{\natexlab{a}})},\ \Eprint
  {http://arxiv.org/abs/arxiv/1006.2081} {arxiv/1006.2081} \BibitemShut
  {NoStop}%
\bibitem [{\citenamefont {Liao}\ and\ \citenamefont
  {Hsiau}(2010)}]{LiaoHsiau2010}%
  \BibitemOpen
  \bibfield  {author} {\bibinfo {author} {\bibfnamefont {C.-C.}\ \bibnamefont
  {Liao}}\ and\ \bibinfo {author} {\bibfnamefont {S.-S.}\ \bibnamefont
  {Hsiau}},\ }\href {\doibase 10.1016/j.powtec.2009.09.017} {\bibfield
  {journal} {\bibinfo  {journal} {Powder Technology}\ }\textbf {\bibinfo
  {volume} {197}},\ \bibinfo {pages} {222 } (\bibinfo {year}
  {2010})}\BibitemShut {NoStop}%
\bibitem [{\citenamefont {Remy}\ \emph {et~al.}(2012)\citenamefont {Remy},
  \citenamefont {Khinast},\ and\ \citenamefont
  {Glasser}}]{RemyKhinastGlasser2012}%
  \BibitemOpen
  \bibfield  {author} {\bibinfo {author} {\bibfnamefont {B.}~\bibnamefont
  {Remy}}, \bibinfo {author} {\bibfnamefont {J.~G.}\ \bibnamefont {Khinast}}, \
  and\ \bibinfo {author} {\bibfnamefont {B.~J.}\ \bibnamefont {Glasser}},\
  }\href {\doibase 10.1002/aic.13743} {\bibfield  {journal} {\bibinfo
  {journal} {AIChE Journal}\ }\textbf {\bibinfo {volume} {to appear}} (\bibinfo
  {year} {2012}),\ 10.1002/aic.13743}\BibitemShut {NoStop}%
\bibitem [{\citenamefont {Herminghaus}(2005)}]{Herminghaus05}%
  \BibitemOpen
  \bibfield  {author} {\bibinfo {author} {\bibfnamefont {S.}~\bibnamefont
  {Herminghaus}},\ }\href@noop {} {\bibfield  {journal} {\bibinfo  {journal}
  {Adv. in Phys.}\ }\textbf {\bibinfo {volume} {54}},\ \bibinfo {pages} {221}
  (\bibinfo {year} {2005})}\BibitemShut {NoStop}%
\bibitem [{\citenamefont {Mitarai}\ and\ \citenamefont
  {Nori}(2006)}]{MitaraiNori2006}%
  \BibitemOpen
  \bibfield  {author} {\bibinfo {author} {\bibfnamefont {N.}~\bibnamefont
  {Mitarai}}\ and\ \bibinfo {author} {\bibfnamefont {F.}~\bibnamefont {Nori}},\
  }\href {\doibase 10.1080/00018730600626065} {\bibfield  {journal} {\bibinfo
  {journal} {Advances in Physics}\ }\textbf {\bibinfo {volume} {55}},\ \bibinfo
  {pages} {1} (\bibinfo {year} {2006})}\BibitemShut {NoStop}%
\bibitem [{\citenamefont {{Ebrahimnazhad Rahbari}}\ \emph
  {et~al.}(2009)\citenamefont {{Ebrahimnazhad Rahbari}}, \citenamefont
  {Vollmer}, \citenamefont {Herminghaus},\ and\ \citenamefont
  {Brinkmann}}]{RahbariVollmerHerminghausBrinkmann2009}%
  \BibitemOpen
  \bibfield  {author} {\bibinfo {author} {\bibfnamefont {S.~H.}\ \bibnamefont
  {{Ebrahimnazhad Rahbari}}}, \bibinfo {author} {\bibfnamefont
  {J.}~\bibnamefont {Vollmer}}, \bibinfo {author} {\bibfnamefont
  {S.}~\bibnamefont {Herminghaus}}, \ and\ \bibinfo {author} {\bibfnamefont
  {M.}~\bibnamefont {Brinkmann}},\ }\href {\doibase 10.1209/0295-5075/87/14002}
  {\bibfield  {journal} {\bibinfo  {journal} {Europhys. Lett.}\ }\textbf
  {\bibinfo {volume} {87}},\ \bibinfo {pages} {14002} (\bibinfo {year}
  {2009})}\BibitemShut {NoStop}%
\bibitem [{\citenamefont {Schulz}\ \emph {et~al.}(2003)\citenamefont {Schulz},
  \citenamefont {Schulz},\ and\ \citenamefont {Herminghaus}}]{Schulz03}%
  \BibitemOpen
  \bibfield  {author} {\bibinfo {author} {\bibfnamefont {M.}~\bibnamefont
  {Schulz}}, \bibinfo {author} {\bibfnamefont {B.~M.}\ \bibnamefont {Schulz}},
  \ and\ \bibinfo {author} {\bibfnamefont {S.}~\bibnamefont {Herminghaus}},\
  }\href {\doibase 10.1103/PhysRevE.67.052301} {\bibfield  {journal} {\bibinfo
  {journal} {Phys. Rev. E}\ }\textbf {\bibinfo {volume} {67}},\ \bibinfo
  {pages} {052301} (\bibinfo {year} {2003})}\BibitemShut {NoStop}%
\bibitem [{\citenamefont {Roeller}\ \emph {et~al.}(2009)\citenamefont
  {Roeller}, \citenamefont {Vollmer},\ and\ \citenamefont
  {Herminghaus}}]{roeller2009}%
  \BibitemOpen
  \bibfield  {author} {\bibinfo {author} {\bibfnamefont {K.}~\bibnamefont
  {Roeller}}, \bibinfo {author} {\bibfnamefont {J.}~\bibnamefont {Vollmer}}, \
  and\ \bibinfo {author} {\bibfnamefont {S.}~\bibnamefont {Herminghaus}},\
  }\href {\doibase 10.1063/1.3202616} {\bibfield  {journal} {\bibinfo
  {journal} {Chaos: An Interdisciplinary Journal of Nonlinear Science}\
  }\textbf {\bibinfo {volume} {19}},\ \bibinfo {eid} {041106} (\bibinfo {year}
  {2009})}\BibitemShut {NoStop}%
\bibitem [{\citenamefont {Rahbari}\ \emph {et~al.}(2010)\citenamefont
  {Rahbari}, \citenamefont {Vollmer}, \citenamefont {Herminghaus},\ and\
  \citenamefont {Brinkmann}}]{vollmer2010}%
  \BibitemOpen
  \bibfield  {author} {\bibinfo {author} {\bibfnamefont {S.~H.~E.}\
  \bibnamefont {Rahbari}}, \bibinfo {author} {\bibfnamefont {J.}~\bibnamefont
  {Vollmer}}, \bibinfo {author} {\bibfnamefont {S.}~\bibnamefont
  {Herminghaus}}, \ and\ \bibinfo {author} {\bibfnamefont {M.}~\bibnamefont
  {Brinkmann}},\ }\href {\doibase 10.1103/PhysRevE.82.061305} {\bibfield
  {journal} {\bibinfo  {journal} {Phys. Rev. E}\ }\textbf {\bibinfo {volume}
  {82}},\ \bibinfo {pages} {061305} (\bibinfo {year} {2010})}\BibitemShut
  {NoStop}%
\bibitem [{\citenamefont {Cates}\ \emph {et~al.}(1998)\citenamefont {Cates},
  \citenamefont {Wittmer}, \citenamefont {Bouchaud},\ and\ \citenamefont
  {Claudin}}]{CatesWittmerBouchaudClaudin1998}%
  \BibitemOpen
  \bibfield  {author} {\bibinfo {author} {\bibfnamefont {M.~E.}\ \bibnamefont
  {Cates}}, \bibinfo {author} {\bibfnamefont {J.~P.}\ \bibnamefont {Wittmer}},
  \bibinfo {author} {\bibfnamefont {J.-P.}\ \bibnamefont {Bouchaud}}, \ and\
  \bibinfo {author} {\bibfnamefont {P.}~\bibnamefont {Claudin}},\ }\href@noop
  {} {\bibfield  {journal} {\bibinfo  {journal} {Phys. Rev. Lett.}\ }\textbf
  {\bibinfo {volume} {81}},\ \bibinfo {pages} {1841} (\bibinfo {year}
  {1998})}\BibitemShut {NoStop}%
\bibitem [{\citenamefont {O'Hern}\ \emph {et~al.}(2003)\citenamefont {O'Hern},
  \citenamefont {Silbert}, \citenamefont {Liu},\ and\ \citenamefont
  {Nagel}}]{OHernSilbertLiuNagel2003}%
  \BibitemOpen
  \bibfield  {author} {\bibinfo {author} {\bibfnamefont {C.~S.}\ \bibnamefont
  {O'Hern}}, \bibinfo {author} {\bibfnamefont {L.~E.}\ \bibnamefont {Silbert}},
  \bibinfo {author} {\bibfnamefont {A.~J.}\ \bibnamefont {Liu}}, \ and\
  \bibinfo {author} {\bibfnamefont {S.~R.}\ \bibnamefont {Nagel}},\ }\href
  {\doibase 10.1103/PhysRevE.68.011306} {\bibfield  {journal} {\bibinfo
  {journal} {Phys. Rev. E}\ }\textbf {\bibinfo {volume} {68}},\ \bibinfo
  {pages} {011306} (\bibinfo {year} {2003})}\BibitemShut {NoStop}%
\bibitem [{\citenamefont {Drocco}\ \emph {et~al.}(2005)\citenamefont {Drocco},
  \citenamefont {Hastings}, \citenamefont {{Olson Reichhardt}},\ and\
  \citenamefont {Reichhardt}}]{DroccoHastingsReichhardtReichhardt2005}%
  \BibitemOpen
  \bibfield  {author} {\bibinfo {author} {\bibfnamefont {J.~A.}\ \bibnamefont
  {Drocco}}, \bibinfo {author} {\bibfnamefont {M.~B.}\ \bibnamefont
  {Hastings}}, \bibinfo {author} {\bibfnamefont {C.~J.}\ \bibnamefont {{Olson
  Reichhardt}}}, \ and\ \bibinfo {author} {\bibfnamefont {C.}~\bibnamefont
  {Reichhardt}},\ }\href@noop {} {\bibfield  {journal} {\bibinfo  {journal}
  {Phys. Rev. Lett.}\ }\textbf {\bibinfo {volume} {95}},\ \bibinfo {pages}
  {088001} (\bibinfo {year} {2005})}\BibitemShut {NoStop}%
\bibitem [{\citenamefont {Valverde}\ \emph {et~al.}(2004)\citenamefont
  {Valverde}, \citenamefont {Quintanilla},\ and\ \citenamefont
  {Castellanos}}]{Valverde2004}%
  \BibitemOpen
  \bibfield  {author} {\bibinfo {author} {\bibfnamefont {J.~M.}\ \bibnamefont
  {Valverde}}, \bibinfo {author} {\bibfnamefont {M.~A.~S.}\ \bibnamefont
  {Quintanilla}}, \ and\ \bibinfo {author} {\bibfnamefont {A.}~\bibnamefont
  {Castellanos}},\ }\href {\doibase 10.1103/PhysRevLett.92.258303} {\bibfield
  {journal} {\bibinfo  {journal} {Phys. Rev. Lett.}\ }\textbf {\bibinfo
  {volume} {92}},\ \bibinfo {pages} {258303} (\bibinfo {year}
  {2004})}\BibitemShut {NoStop}%
\bibitem [{\citenamefont {Fingerle}\ \emph {et~al.}(2008)\citenamefont
  {Fingerle}, \citenamefont {Roeller}, \citenamefont {Huang},\ and\
  \citenamefont {Herminghaus}}]{Fingerle08}%
  \BibitemOpen
  \bibfield  {author} {\bibinfo {author} {\bibfnamefont {A.}~\bibnamefont
  {Fingerle}}, \bibinfo {author} {\bibfnamefont {K.}~\bibnamefont {Roeller}},
  \bibinfo {author} {\bibfnamefont {K.}~\bibnamefont {Huang}}, \ and\ \bibinfo
  {author} {\bibfnamefont {S.}~\bibnamefont {Herminghaus}},\ }\href@noop {}
  {\bibfield  {journal} {\bibinfo  {journal} {New Journal of Physics}\ }\textbf
  {\bibinfo {volume} {10}},\ \bibinfo {pages} {053020 (10pp)} (\bibinfo {year}
  {2008})}\BibitemShut {NoStop}%
\bibitem [{\citenamefont {Huang}\ \emph {et~al.}(2009)\citenamefont {Huang},
  \citenamefont {Roeller},\ and\ \citenamefont {Herminghaus}}]{HuangRoeller09}%
  \BibitemOpen
  \bibfield  {author} {\bibinfo {author} {\bibfnamefont {K.}~\bibnamefont
  {Huang}}, \bibinfo {author} {\bibfnamefont {K.}~\bibnamefont {Roeller}}, \
  and\ \bibinfo {author} {\bibfnamefont {S.}~\bibnamefont {Herminghaus}},\
  }\href@noop {} {\bibfield  {journal} {\bibinfo  {journal} {The European
  Physical Journal - Special Topics}\ }\textbf {\bibinfo {volume} {179}},\
  \bibinfo {pages} {25} (\bibinfo {year} {2009})},\ \bibinfo {note}
  {10.1140/epjst/e2010-01191-5}\BibitemShut {NoStop}%
\bibitem [{\citenamefont {Ulrich}\ \emph
  {et~al.}(2009{\natexlab{a}})\citenamefont {Ulrich}, \citenamefont
  {Aspelmeier}, \citenamefont {Roeller}, \citenamefont {Fingerle},
  \citenamefont {Herminghaus},\ and\ \citenamefont {Zippelius}}]{ulrich09}%
  \BibitemOpen
  \bibfield  {author} {\bibinfo {author} {\bibfnamefont {S.}~\bibnamefont
  {Ulrich}}, \bibinfo {author} {\bibfnamefont {T.}~\bibnamefont {Aspelmeier}},
  \bibinfo {author} {\bibfnamefont {K.}~\bibnamefont {Roeller}}, \bibinfo
  {author} {\bibfnamefont {A.}~\bibnamefont {Fingerle}}, \bibinfo {author}
  {\bibfnamefont {S.}~\bibnamefont {Herminghaus}}, \ and\ \bibinfo {author}
  {\bibfnamefont {A.}~\bibnamefont {Zippelius}},\ }\href {\doibase
  10.1103/PhysRevLett.102.148002} {\bibfield  {journal} {\bibinfo  {journal}
  {Physical Review Letters}\ }\textbf {\bibinfo {volume} {102}},\ \bibinfo
  {eid} {148002} (\bibinfo {year} {2009}{\natexlab{a}})}\BibitemShut {NoStop}%
\bibitem [{\citenamefont {Ulrich}\ \emph
  {et~al.}(2009{\natexlab{b}})\citenamefont {Ulrich}, \citenamefont
  {Aspelmeier}, \citenamefont {Zippelius}, \citenamefont {Roeller},
  \citenamefont {Fingerle},\ and\ \citenamefont {Herminghaus}}]{ulrich09_PRE}%
  \BibitemOpen
  \bibfield  {author} {\bibinfo {author} {\bibfnamefont {S.}~\bibnamefont
  {Ulrich}}, \bibinfo {author} {\bibfnamefont {T.}~\bibnamefont {Aspelmeier}},
  \bibinfo {author} {\bibfnamefont {A.}~\bibnamefont {Zippelius}}, \bibinfo
  {author} {\bibfnamefont {K.}~\bibnamefont {Roeller}}, \bibinfo {author}
  {\bibfnamefont {A.}~\bibnamefont {Fingerle}}, \ and\ \bibinfo {author}
  {\bibfnamefont {S.}~\bibnamefont {Herminghaus}},\ }\href {\doibase
  10.1103/PhysRevE.80.031306} {\bibfield  {journal} {\bibinfo  {journal}
  {Physical Review E (Statistical, Nonlinear, and Soft Matter Physics)}\
  }\textbf {\bibinfo {volume} {80}},\ \bibinfo {eid} {031306} (\bibinfo {year}
  {2009}{\natexlab{b}})}\BibitemShut {NoStop}%
\bibitem [{\citenamefont {Roeller}(2010)}]{ThesisKlaus10}%
  \BibitemOpen
  \bibfield  {author} {\bibinfo {author} {\bibfnamefont {K.}~\bibnamefont
  {Roeller}},\ }\emph {\bibinfo {title} {Numerical simulations of wet granular
  matter}},\ \href@noop {} {Ph.D. thesis},\ \bibinfo  {school} {University
  Goettingen} (\bibinfo {year} {2010})\BibitemShut {NoStop}%
\bibitem [{\citenamefont {Davis}\ \emph {et~al.}(2002)\citenamefont {Davis},
  \citenamefont {Rager},\ and\ \citenamefont {Good}}]{DavisRagerGood2002}%
  \BibitemOpen
  \bibfield  {author} {\bibinfo {author} {\bibfnamefont {R.~H.}\ \bibnamefont
  {Davis}}, \bibinfo {author} {\bibfnamefont {D.~A.}\ \bibnamefont {Rager}}, \
  and\ \bibinfo {author} {\bibfnamefont {B.~T.}\ \bibnamefont {Good}},\ }\href
  {\doibase 10.1017/S0022112002001489} {\bibfield  {journal} {\bibinfo
  {journal} {Journal of Fluid Mechanics}\ }\textbf {\bibinfo {volume} {468}},\
  \bibinfo {pages} {107 } (\bibinfo {year} {2002})}\BibitemShut {NoStop}%
\bibitem [{\citenamefont {Antonyuk}\ \emph {et~al.}(2009)\citenamefont
  {Antonyuk}, \citenamefont {Heinrich}, \citenamefont {Deen},\ and\
  \citenamefont {Kuipers}}]{AntonyukHeinrichDeenKuipers2009}%
  \BibitemOpen
  \bibfield  {author} {\bibinfo {author} {\bibfnamefont {S.}~\bibnamefont
  {Antonyuk}}, \bibinfo {author} {\bibfnamefont {S.}~\bibnamefont {Heinrich}},
  \bibinfo {author} {\bibfnamefont {N.}~\bibnamefont {Deen}}, \ and\ \bibinfo
  {author} {\bibfnamefont {H.}~\bibnamefont {Kuipers}},\ }\href {\doibase
  10.1016/j.partic.2009.04.006} {\bibfield  {journal} {\bibinfo  {journal}
  {Particuology}\ }\textbf {\bibinfo {volume} {7}},\ \bibinfo {pages} {245 }
  (\bibinfo {year} {2009})},\ \bibinfo {note} {from micro-scale to technical
  dimension - Challenges in the simulation of dense gas-particle flows,
  Selected papers from the Sino-German Workshop.}\BibitemShut {Stop}%
\bibitem [{\citenamefont {Donahue}\ \emph
  {et~al.}(2010{\natexlab{b}})\citenamefont {Donahue}, \citenamefont {Hrenya},
  \citenamefont {Davis}, \citenamefont {Nakagawa}, \citenamefont {Zelinskaya},\
  and\ \citenamefont
  {Joseph}}]{DonahueHrenyaDavisNakagawaZelinskayaJoseph2010}%
  \BibitemOpen
  \bibfield  {author} {\bibinfo {author} {\bibfnamefont {C.~M.}\ \bibnamefont
  {Donahue}}, \bibinfo {author} {\bibfnamefont {C.~M.}\ \bibnamefont {Hrenya}},
  \bibinfo {author} {\bibfnamefont {R.~H.}\ \bibnamefont {Davis}}, \bibinfo
  {author} {\bibfnamefont {K.~J.}\ \bibnamefont {Nakagawa}}, \bibinfo {author}
  {\bibfnamefont {A.~P.}\ \bibnamefont {Zelinskaya}}, \ and\ \bibinfo {author}
  {\bibfnamefont {G.~G.}\ \bibnamefont {Joseph}},\ }\href {\doibase
  10.1017/S0022112009993715} {\bibfield  {journal} {\bibinfo  {journal}
  {Journal of Fluid Mechanics}\ }\textbf {\bibinfo {volume} {650}},\ \bibinfo
  {pages} {479} (\bibinfo {year} {2010}{\natexlab{b}})}\BibitemShut {NoStop}%
\bibitem [{\citenamefont {Donahue}\ \emph
  {et~al.}(2012{\natexlab{a}})\citenamefont {Donahue}, \citenamefont {Brewer},
  \citenamefont {Davis},\ and\ \citenamefont
  {Hrenya}}]{DonahueBrewerDavisHrenya2012}%
  \BibitemOpen
  \bibfield  {author} {\bibinfo {author} {\bibfnamefont {C.~M.}\ \bibnamefont
  {Donahue}}, \bibinfo {author} {\bibfnamefont {W.~M.}\ \bibnamefont {Brewer}},
  \bibinfo {author} {\bibfnamefont {R.~H.}\ \bibnamefont {Davis}}, \ and\
  \bibinfo {author} {\bibfnamefont {C.~M.}\ \bibnamefont {Hrenya}},\ }\href
  {\doibase 10.1017/jfm.2012.297} {\bibfield  {journal} {\bibinfo  {journal}
  {Journal of Fluid Mechanics}\ }\textbf {\bibinfo {volume} {708}},\ \bibinfo
  {pages} {128} (\bibinfo {year} {2012}{\natexlab{a}})}\BibitemShut {NoStop}%
\bibitem [{\citenamefont {Donahue}\ \emph
  {et~al.}(2012{\natexlab{b}})\citenamefont {Donahue}, \citenamefont {Davis},
  \citenamefont {Kantak},\ and\ \citenamefont
  {Hrenya}}]{DonahueDavisKantakHrenya2012}%
  \BibitemOpen
  \bibfield  {author} {\bibinfo {author} {\bibfnamefont {C.~M.}\ \bibnamefont
  {Donahue}}, \bibinfo {author} {\bibfnamefont {R.~H.}\ \bibnamefont {Davis}},
  \bibinfo {author} {\bibfnamefont {A.~A.}\ \bibnamefont {Kantak}}, \ and\
  \bibinfo {author} {\bibfnamefont {C.~M.}\ \bibnamefont {Hrenya}},\ }\href
  {\doibase 10.1103/PhysRevE.86.021303} {\bibfield  {journal} {\bibinfo
  {journal} {Phys. Rev. E}\ }\textbf {\bibinfo {volume} {86}},\ \bibinfo
  {pages} {021303} (\bibinfo {year} {2012}{\natexlab{b}})}\BibitemShut
  {NoStop}%
\bibitem [{\citenamefont {Gollwitzer}\ \emph {et~al.}(2012)\citenamefont
  {Gollwitzer}, \citenamefont {Rehberg}, \citenamefont {Kruelle},\ and\
  \citenamefont {Huang}}]{GollwitzerRehbergKruelle2012}%
  \BibitemOpen
  \bibfield  {author} {\bibinfo {author} {\bibfnamefont {F.}~\bibnamefont
  {Gollwitzer}}, \bibinfo {author} {\bibfnamefont {I.}~\bibnamefont {Rehberg}},
  \bibinfo {author} {\bibfnamefont {C.~A.}\ \bibnamefont {Kruelle}}, \ and\
  \bibinfo {author} {\bibfnamefont {K.}~\bibnamefont {Huang}},\ }\href
  {\doibase 10.1103/PhysRevE.86.011303} {\bibfield  {journal} {\bibinfo
  {journal} {Phys. Rev. E}\ }\textbf {\bibinfo {volume} {86}},\ \bibinfo
  {pages} {011303} (\bibinfo {year} {2012})}\BibitemShut {NoStop}%
\bibitem [{\citenamefont {Hoover}(1983)}]{Hoover1983}%
  \BibitemOpen
  \bibfield  {author} {\bibinfo {author} {\bibfnamefont {W.~G.}\ \bibnamefont
  {Hoover}},\ }\href {\doibase 10.1146/annurev.pc.34.100183.000535} {\bibfield
  {journal} {\bibinfo  {journal} {Annual Review of Physical Chemistry}\
  }\textbf {\bibinfo {volume} {34}},\ \bibinfo {pages} {103} (\bibinfo {year}
  {1983})}\BibitemShut {NoStop}%
\bibitem [{\citenamefont {Schulz}\ and\ \citenamefont
  {Schulz}(2006)}]{Schulz2006}%
  \BibitemOpen
  \bibfield  {author} {\bibinfo {author} {\bibfnamefont {B.}~\bibnamefont
  {Schulz}}\ and\ \bibinfo {author} {\bibfnamefont {M.}~\bibnamefont
  {Schulz}},\ }\href {\doibase 10.1016/j.jnoncrysol.2006.03.125} {\bibfield
  {journal} {\bibinfo  {journal} {Journal of Non-Crystalline Solids}\ }\textbf
  {\bibinfo {volume} {352}},\ \bibinfo {pages} {4877 } (\bibinfo {year}
  {2006})},\ \bibinfo {note} {proceedings of the 5th International Discussion
  Meeting on Relaxations in Complex Systems, 5th International Discussion
  Meeting on Relaxations in Complex Systems}\BibitemShut {NoStop}%
\bibitem [{\citenamefont {Rahbari}\ \emph {et~al.}(2009)\citenamefont
  {Rahbari}, \citenamefont {Vollmer}, \citenamefont {Herminghaus},\ and\
  \citenamefont {Brinkmann}}]{vollmer09}%
  \BibitemOpen
  \bibfield  {author} {\bibinfo {author} {\bibfnamefont {S.~H.~E.}\
  \bibnamefont {Rahbari}}, \bibinfo {author} {\bibfnamefont {J.}~\bibnamefont
  {Vollmer}}, \bibinfo {author} {\bibfnamefont {S.}~\bibnamefont
  {Herminghaus}}, \ and\ \bibinfo {author} {\bibfnamefont {M.}~\bibnamefont
  {Brinkmann}},\ }\href@noop {} {\bibfield  {journal} {\bibinfo  {journal} {EPL
  (Europhysics Letters)}\ }\textbf {\bibinfo {volume} {87}},\ \bibinfo {pages}
  {14002} (\bibinfo {year} {2009})}\BibitemShut {NoStop}%
\bibitem [{Note1()}]{Note1}%
  \BibitemOpen
  \bibinfo {note} {In \cite {vollmer2010} we rather used the amplitude of the
  velocity profile in the direction of the external field (i.e., the amplitude
  of the velocity response in reaction to the applied force field, \protect
  \ensuremath {F_{s\protect \mathrm {max}}}, as an order parameter. This has
  advantages when following the hysteresis loop of the response upon slowly
  decreasing and subsequently increasing \protect \ensuremath {F_{s\protect
  \mathrm {max}}}. However, in the present study, where we focus on the arrest
  of flow, the kinetic energy, $E_y$, turned out to be a numerically stable and
  easier accessible order parameter.}\BibitemShut {Stop}%
\end{thebibliography}
%

\end{document}